\documentclass[a4paper]{article}

\usepackage{INTERSPEECH2021}
\usepackage{tabularx}
\usepackage{bm}
\usepackage{booktabs}
\usepackage{graphicx}
\usepackage{array}

\title{Cross-linguistic Prosodic Analysis of Autistic and Non-autistic Child Speech in Finnish, French and
Slovak}

\name{Ida-Lotta Myllylä$^1$, Sofoklis Kakouros$^2$}
\address{
  $^1$Department of Languages, University of Helsinki\\
  $^2$Speech and Cognition Research Group, Signal Processing Research Centre, Tampere University}
\email{ida-lotta.myllyla@helsinki.fi,sofoklis.kakouros@tuni.fi}
\begin{document}

\maketitle
\begin{abstract}

%Prosodic differences in autism are well-documented, but cross-linguistic evidence remains limited. This study investigates prosody in autism in a multilingual corpus of Finnish, French, and Slovak speakers. A set of 88 acoustic features were extracted from $>$ 5000 inter-pausal units. PCA was used to reduce the data into orthogonal components. Group differences were analyzed using Linear Mixed-Effects Models.
%Cross-linguistically, autistic speakers exhibited increased general intensity variability and less breathy vocal quality (higher Harmonics-to-Noise Ratio and alpha ratio), but reduced temporal intensity dynamics and lower central $f_0$. Monolingual analyses revealed language-specific nuances: while Slovak results aligned with the cross-linguistic dataset of lower $f_0$ in autism, they diverged on voice quality, whereas Finnish results mirrored the cross-linguistic voice quality patterns. These findings emphasize including voice quality and intensity dynamics in the study of possible language-independent markers of prosody in autism, alongside the traditional focus on pitch. The results challenge simple deficiency-based models, pointing instead to a complex, acoustically distinct prosodic profile characterized by  distinct properties in voice quality and intensity quality across languages.

Prosodic differences in autism are well-documented, but cross-linguistic evidence remains limited. This study investigates prosody in autism across a multilingual corpus of Finnish, French, and Slovak speakers. 88 acoustic features from over 5,000 inter-pausal units were extracted, and data were reduced via Principal Component Analysis (PCA) and analyzed using Linear Mixed-Effects Models (LMMs).

Cross-linguistically, autistic speakers exhibited increased general intensity variability and a clearer, less breathy voice quality (higher Harmonics-to-Noise Ratio and alpha ratio), alongside reduced temporal intensity dynamics and lower central $f_0$. Monolingual analyses revealed language-specific nuances: Slovak results aligned with cross-linguistic $f_0$ patterns but diverged on voice quality, while Finnish results mirrored the broader voice quality findings. 

These results emphasize including voice quality and intensity dynamics in the study of possible language-independent markers of autism, alongside traditional pitch measures. The findings challenge deficiency-based models, suggesting instead a complex, acoustically distinct prosodic profile across languages.

\end{abstract}

\noindent\textbf{Index Terms}: prosody, child speech, autism, Finnish, French, Slovak 

\section{Introduction}

Autism is a neurodevelopmental condition characterized by differences in social communication and interaction \cite{lord2018autism}. Pragmatic communication in autism may diverge from neurotypical norms \cite{grice2023linguistic}, and prosody may be different compared to non-autistic peers \cite{fusaroli2022toward}. Despite significant population heterogeneity and speaker-related differences, pitch (fundamental frequency, $f_0$) has been noted as a primary area of divergence \cite{fusaroli2017voice,fusaroli2022toward}. Acoustic analyses report both increased $f_0$ variability/range and conversely, low variability; such prosodic extremes may be salient markers in autism  \cite{wiklund2019, fusaroli2022toward}. Other features include diverging intensity, temporal prosody, and voice quality \cite{trayvick2024speech, fusaroli2017voice, fusaroli2022toward,guo2022applying}. While autism-adjacent prosodic features have been identified within and across languages \cite{fusaroli2022toward,lau2022cross}, cross-linguistic evidence remains limited. Meta-analyses highlight $f_0$ as a robust marker, but voice quality features lack empirical clarity \cite{trayvick2024speech, guo2022applying}. We address these gaps by analyzing multi- and monolingual datasets to isolate prosodic features possibly differentiating diagnostic groups.

Adopting a neurodiversity framework, we view autism as a condition rather than a disorder. We utilize identity-first language ('autistic speakers') to reflect the preferences of the autistic community \cite{bottema2021avoiding}. The diagnostic term Autism Spectrum Disorder (ASD) is retained when referring to clinical classification and to ensure comparability with literature. We emphasize that prosodic differences, while 'atypical' by neurotypical standards, are not inherent deficits. Rather, they may stem from distinct neuromotor characteristics or alternative communicative priorities, such as highlighting semantic information or regulating cognitive load \cite{diehl2013acoustic, thurber1993pauses}. Thus, these patterns may represent functional adaptations to unique sensory and cognitive processing styles in autism \cite{fusaroli2017voice, milton2012ontological}.

%This template can be found on the conference website. %Templates are provided for Microsoft Word\textregistered, and \LaTeX. However, we highly recommend using \LaTeX when preparing your submission. Information for full paper submission is available on the conference website.

\section{Materials}
As summarized in Table \ref{tab:data_summary}, three datasets were used, one for each language under study: Finnish, French, and Slovak. 

\subsection{Finnish dataset}
Finnish data were collected in Helsinki, Finland from six native Finnish-speaking autistic males (aged 11–13) diagnosed with Asperger’s syndrome (ICD-10 \cite{ICD10_2004}). The control data were collected from six age-matched, non-autistic males. The autistic participants were recorded during spontaneous group discussions in a hospital-based social skills intervention. Each session lasted approximately two hours, comprising free-form discussion and board games. Three participants attended one session, and three another. The control group was recorded in a comparable teacher-led free-form discussion at a school. All sessions were recorded in quiet environments using portable recorders and head-mounted microphones (16-bit mono, 44.1 kHz).

%The control (CRT) data were collected from six age-matched, non-autistic, native Finnish-speaking males with no reported developmental concerns. The autistic participants were recorded during spontaneous group discussions in a hospital-based social skills and communication intervention in the Helsinki metropolitan area. Three participants attended one session, and three another. Each session lasted approximately two hours, comprising free-form discussion and games. The sessions were led by two healthcare professionals. The control group was recorded in a comparable teacher-led discussion at a Helsinki metropolitan area school. This conversational setting was constructed for research purposes, consisting of free-form discussion. All participants were selected based on availability from their respective locations. All sessions were recorded in quiet environments using portable recorders and head-mounted microphones for each speaker, capturing 16-bit mono audio at 44.1 kHz. 

\subsection{French dataset}
French data were collected in Geneva and Lausanne, Switzerland from six native French-speaking autistic males and three non-autistic controls (aged 11–13). Autistic participants were recorded during speech therapy group interventions. Each session lasted approximately an hour, comprising free-form discussion and games. Two participants attended one session, and four another. The control group was recorded in a comparable adult-led free-form discussion setting. All sessions were recorded in quiet environments using portable recorders and head-mounted microphones (16-bit mono, 44.1 kHz).

%French speech data were collected from six native French-speaking autistic males and three non-autistic controls, all aged 11–14. The autistic participants were recorded during spontaneous group sessions, a speech therapy intervention for communication. Two participants attended one session, and four another. Each session lasted approximately an hour, comprising free-form discussion and games. Sessions were led by two speech therapists. The session with three speakers was recorded in Geneva, Switzerland and the other in Orbe, Switzerland. The control group was recorded in an adult-led discussion in Lausanne, Switzerland, a conversational setting constructed for research purposes, which comprised free-form discussion. All participants groups were selected based on availability from their respective locations. All sessions were recorded in quiet environments using portable recorders and head-mounted microphones for each speaker, capturing 16-bit mono audio at 44.1 kHz. 

\subsection{Slovak dataset}
Slovak data utilized is the Slovak Autistic and Non-Autistic Child Speech Corpus (SANACS) \cite{kruyt2024slovak}, consisting of collaborative, task-oriented conversations between children diagnosed with ASD and a non-autistic adult experimenter. The data entail speech from 67 children aged 6-12 (mean 9.2): 37 autistic (7 female, 30 male) and 30 non-autistic (7 female, 23 male). The Maps task was used in elicitation, with each child acting both as a follower and describer during the task. All sessions were recorded with portable recorders and microphones capturing stereo audio, from which the child speech tracks were extracted. The corpus was recorded by the Institute of Informatics, Slovak Academy of Sciences, and the Academic Research Center for Autism, at the Comenius University in Bratislava.

\begin{table}[t]
\centering
\small
\setlength{\tabcolsep}{3pt}
\resizebox{\columnwidth}{!}{%
  \begin{tabular}{lcccc}
    \toprule
    Language & ASD (IPUs/IDs) & CRT (IPUs/IDs) & Mean dur. (s) & Total (IPUs/IDs) \\
    \midrule
    Finnish & 1323 / 6  & 249 / 6   & 2.11 & 1572 / 12 \\
    French  & 884 / 6   & 659 / 3   & 1.79 & 1543 / 9  \\
    Slovak  & 2867 / 37 & 2230 / 29 & 2.36 & 5097 / 66 \\
    \midrule
    ALL     & 5074 / 49 & 3138 / 38 & 2.20 & 8212 / 87 \\
    \bottomrule
  \end{tabular}
}
\caption{Summary of IPUs, unique speaker IDs per diagnostic group, and mean recording duration by language.}
\label{tab:data_summary}
\vspace{-2.0\baselineskip}
\end{table}

\subsection{Data preprocessing}
All speech data were segmented into IPUs (inter-pausal units), i.e. sections of speech between pauses of at least 200 ms, produced by one speaker. The segmenting was done automatically using intensity thresholds and validated manually. IPUs with overlapping speech and those consisting only of non-lexical vocalizations, were discarded prior to the analysis. One speaker from the Slovak dataset was excluded at this stage due to a substantial amount of overlapping speech and IPUs consisting only of non-lexical hesitation markers.

\section{Methods}

\subsection{Feature extraction}
All speech recordings were processed with openSMILE, an open-source toolkit for audio feature extraction, using the eGeMAPS utterance-level functionals configuration \cite{eyben2015geneva}, producing an 88-dimensional feature vector for each IPU. This configuration combines selected low-level descriptors with their statistical functionals, covering $f_0$, intensity, spectral characteristics, and voice quality measures. The extracted features provide a comprehensive representation of prosodic dynamics and voice quality characteristics capturing pitch and energy distribution, spectral balance, and temporal variability, capturing a standardized, physiologically relevant baseline for phonetic research. While not exhaustive, this configuration ensures comparability with existing literature while isolating key markers for future high-dimensional exploration. Feature outputs were compiled to form the cross-linguistic dataset, which included metadata for each IPU (i.e., speaker ID, group, language, sex and age) needed for statistical modeling. Per-language datasets were compiled accordingly.

%(crosslinguistic, Finnish, French, and Slovak). 

\subsection{Statistical analysis}
\subsubsection{Principal Component Analysis}

To address the multicollinearity in the extracted features, Principal Component Analysis (PCA) was employed for all datasets. PCA is a dimensionality reduction technique that transforms a set of correlated variables into a smaller set of uncorrelated components, each capturing a distinct dimension of variation in the data. Each component is defined by loadings, the correlations between the original features and that component, which reflect how strongly each feature contributes to the component and allow for acoustic interpretation of what each component represents. Prior to analysis, outliers were removed from the data using Winsorization, clamping values at the 2.5th and 97.5th percentiles. This data was then z-scored to ensure proportionate influence of all features in the analysis.

%with large values (e.g., $f_0$) did not disproportionately influence the analysis over features with small values (e.g., jitter).

PCA was performed in two stages: first on the combined cross-linguistic dataset, and then separately on each per-language dataset, enabling the identification of prosodic patterns across and within languages. PCA was performed on the scaled 88-feature matrices, and the first 10 principal components from each dataset were retained for analysis based on two criteria: 1) all 10 components exhibited a variance (eigenvalue) greater than 1.0, satisfying Kaiser's rule, and 2) these 10 components collectively explained a substantial percentage of the cumulative variance (cross-linguistic dataset 63.69\%; Finnish: 60.33\%; French: 65.48\%; Slovak: 65.43\%). This robust compromise significantly reduced data dimensionality while retaining maximum acoustic prosodic variability. The sources of remaining variance were not explored further, as they may stem from fine-grained idiosyncratic or situational differences, a research focus we chose to leave for future investigations. 

The components for each dataset were identified by interpreting the top 10 features with the highest absolute loadings within each PC. To ensure non-redundant definitions, a "maximum loading rule" was applied: if a feature appeared in the top 10 of multiple PCs, it was assigned only to the component in which it had the highest absolute loading, guaranteeing that each defining feature contributed maximally to the concept represented by its assigned PC. These orthogonal components were used in subsequent mixed-effects modeling (see PC summaries in Tables 2—4).

While PCA revealed component structures consistent with the other languages (e.g., distinct intensity, $f_0$, and voice quality dimensions) in the French dataset, subsequent modeling yielded no significant differences due to the limited sample size. Consequently, neither the PCs nor the mixed-effects modeling are not detailed further. However, the French data contributed effectively to the pooled cross-linguistic model, aiding the identification of shared prosodic patterns independent of language-specific baselines.

\begin{table}[h]
\centering
\setlength{\tabcolsep}{3pt}
\scriptsize %
\label{tab:placeholder_label}
\begin{tabularx}{\columnwidth}{l X}
\toprule 
PC & Description \\ 
\midrule 
PC1 & Overall and Central Intensity \\ 
PC2 & Spectral Balance, MFCC-1 \\ 
PC3 & Intensity Dynamics, Central $f_0$, Spectral Change \\ 
PC4 & Voice Clarity, Harmonicity \\ 
PC5 & $f_0$ Variation, Rapid $f_0$ Perturbation (Jitter) \\ 
PC6 & Formant Resonance, Bandwidth Variation, Voicing \\ 
PC7 & Unvoiced Spectral Quality, Intensity Consistency \\
PC8 & Formant Frequency and Stability \\
PC9 & Voiced Segment Duration, F1 Stability \\
PC10 & Short-Term Perturbation, Articulation Rate \\ 
\bottomrule 
\end{tabularx}
\caption{Summary of Cross-linguistic Principal Components.}
    \label{tab:crossling_pcs}
\vspace{-2.2em}
\end{table}

\begin{table}[h]
\centering
\setlength{\tabcolsep}{3pt}
\scriptsize %
\label{tab:placeholder_label}
\begin{tabularx}{\columnwidth}{l X}
\toprule 
PC & Description \\ 
\midrule 
PC1 & Overall and Central Intensity \\ 
PC2 & Spectral Balance, Voice Quality \\ 
PC3 & Formants, $f_0$, Clarity \\ 
PC4 & $f_0$ Variation, Spectral Dynamics \\ 
PC5 & $f_0$ Range, Intensity Minimum, Spectral Variation  \\ 
PC6 & F2 Profile, Cepstral Characteristics \\ 
PC7 & Spectral Tilt, $f_0$ Dynamics \\
PC8 & Unvoiced Segment Duration, Shimmer \\
PC9 & Voiced Segment Duration, Jitter \\
PC10 & $f_0$ Falling Dynamics, Acoustic Variability \\ 
\bottomrule 
\end{tabularx}
\caption{Summary of Finnish Principal Components.}
    \label{tab:finnish_pcs}
\vspace{-2.2em}
\end{table}

\begin{table}[h]
\centering
\setlength{\tabcolsep}{3pt}
\scriptsize %
\label{tab:placeholder_label}
\begin{tabularx}{\columnwidth}{l X}
\toprule 
PC & Description \\ 
\midrule 
PC1 & Overall and Dynamic Intensity \\ 
PC2 & Spectral Balance, MFCC-1, Voice Quality \\ 
PC3 & Central $f_0$, Voice Clarity, Articulation Rate \\ 
PC4 & Voice Clarity, MFCC-2, Formant Frequencies \\ 
PC5 & Unvoiced Segment Duration, Acoustic Variability \\ 
PC6 & $f_0$ Variability, Jitter, Shimmer \\ 
PC7 & Formant Stability, Voiced Segment Duration\\
PC8 & Temporal Rhythm, Spectral Tilt \\
PC9 & Unvoiced Spectral Quality, MFCC-4\\
PC10 & $f_0$ Contour Dynamics and Perturbation \\ 
\bottomrule 
\end{tabularx}
\caption{Summary of Slovak Principal Components.}
    \label{tab:slovak_pcs}
\vspace{-2.2em}
\end{table}

\subsubsection{Linear mixed-effects modeling}
For each dataset, we fit separate Linear Mixed-Effects Models (LMMs) for the 10 PCs. The PC score served as the dependent variable, with Group (reference: ASD) as the fixed effect, with standardized age and sex as covariates.

The monolingual models included a random intercept for speaker. The cross-linguistic model employed a nested random intercept structure (speaker within language) to account for hierarchical sources of non-independence. This explicitly models baseline prosodic variations across languages and speakers, ensuring that results reflect group differences regardless of the specific language spoken.

To account for multiple testing, p-values for the group effect were adjusted within modeling using the Benjamini-Hochberg False Discovery Rate (FDR) procedure \cite{benjamini1995controlling}, with a significance threshold of $p < 0.05$. Confidence intervals (95\% CIs) were computed using the profile likelihood method. Finally, for all models, the inspected adjusted Intraclass Correlation Coefficients (ICC) indicated the need for the mixed-effects structure, and model assumptions (homogeneity of variance, normality of residuals and random effects) were confirmed via visual inspection.

\section{Results}
\subsection{Cross-linguistic group comparisons}
In determining how prosodic features differ between speaker groups across languages, the LMMs yielded significant results in three PCs,  comprising different acoustic features: PC3 (Intensity Dynamics, Central $f_0$, Spectral Change), PC4 (Voice Clarity, Harmonicity) and PC7 (Unvoiced Spectral Quality, Intensity Consistency). Results of the LMMs are found in Table 5. A negative estimate indicates a lower score in the control group, and higher score in the autistic speaker group. 

%The adjusted Intraclass Correlation Coefficients (ICC) for all models indicated the need for the nested random effect modeling structure. Visual inspections of the models' diagnostic plots confirmed that model assumptions were met.

PC3 had the top loading features of mean spectral flux as frame-to-frame spectral change, mean $f_0$ in semitones, 20th percentile and median of $f_0$ in semitones, mean smoothed slope of both rising and falling parts of the intensity contour, standard deviation of the slope of falling intensity contour, and mean spectral slope in 0–500Hz band on voiced frames. This PC captured the dynamic intensity changes, central $f_0$ level, and the short-time changes of the spectral qualities. The LMM revealed that the autistic speakers scored significantly lower on component PC3 compared to control speakers, suggesting that the autistic speakers' speech was characterized by lower overall $f_0$, decreased dynamic intensity changes (decreased slopes) and decreased frame-to-frame spectral change, pointing to consistent energy distribution across the spectrum staying relatively constant between frames. 

PC4, a component of voice quality, had top loadings of the mean and standard deviation of Harmonics-to-Noise Ratio (HNR), mean alpha ratio of voiced frames and H1-H2 energy difference. A high score on this PC is associated with a clearer, more powerful, and less breathy voice. The LMM revealed that the autistic speakers scored significantly higher on this component compared to the control speakers. This suggests a profile of either reduced voicing regularity or a different harmonic energy balance in the control speakers, and increased clarity and reduced breathiness in the autistic speaker group. 

PC7 combined voice quality and intensity features, with top loadings of mean alpha ratio and mean Hammarberg index of unvoiced frames, coefficient of variation of intensity (std/mean), and coefficient of variation of spectral flux, and both mean and coefficient of variation of spectral slope in 500–1500Hz band. A high score on this PC points to a specific unvoiced sound quality and less consistent intensity. The LMM revealed the autistic speakers scored significantly higher on this component compared to the control speakers. This suggests a distinct difference in the unvoiced spectral profiles between groups and intensity. Control speakers exhibited less intensity variability and a less acoustically prominent spectral profile in unvoiced segments. Conversely, the autistic speakers exhibited more variable intensity and unvoiced segments with greater high-frequency energy, suggesting an acoustically distinct articulation.

\begin{table}[ht]
    \centering
    \scriptsize %
    \label{tab:lmm_results}
    \begin{tabular}{lrrrrc c}
        \toprule
        \textbf{PC} & $\bm{\beta}$ & \textbf{\textit{SE}} & \textbf{\textit{df}} & \textbf{\textit{t}} & \textbf{FDR-\textit{p}} & \textbf{95\% CI} \\
        \midrule
        PC3 & 1.64 & 0.48 & 79.01 & 3.442 & 0.005 *& [0.72, 1.16] \\
        
        PC4 & -0.86 & 0.31 & 77.02 & -2.71 &  0.03 *& [-1.47, -0.25] \\
        
        PC7 & -0.65 & 0.15 & 79.26 & -4.42 & $<$ 0.01 *& [-0.94, -0.36] \\
        
        % Add more rows as needed...
        \bottomrule
    \end{tabular}
    \caption{Results of the LMMs in the cross-linguistic analysis.}

    \vspace{-2.9em}
\end{table}

\subsection{Finnish group comparisons}
Following the same PC determination process in Finnish, the variety of acoustic features compiled into a unique set of PCs, which were used in subsequent LMMs. Only the model for PC2 (Spectral Balance, Voice Quality) revealed a significant effect. PC2 captured the spectral shape and voice quality features, mean alpha ratio and mean Hammarberg index of voiced frames, mean and variability of MFCC-1, HI-A3 difference, and F1 amplitude and bandwith. 

The LMM of PC2 yielded a significant effect of group ($\beta$ = 3.98, \textit{SE} = 1.05, \textit{df} = 8.51, \textit{t} = 3.80, FDR-adjusted \textit{p} = 0.04, 95\% CI [2.06, 5.90]), indicating that autistic speakers scored lower and control speakers higher on this component defined by spectral balance features. Acoustically, high values in these features correspond to a steeper spectral tilt, where energy is concentrated in the $f_0$ and drops off rapidly in the higher harmonics, a profile characteristic of a breathier phonation style. Consequently, the non-autistic Finnish speakers exhibited a breathier voice quality with greater low-frequency dominance, whereas the autistic speakers displayed a comparatively clearer voice quality, aligning with the voice clarity patterns observed in the cross-linguistic analysis.

\subsection{Slovak group comparisons}
The Slovak-specific analysis was conducted in similar fashion. 
%The adjusted ICCs for LMMs indicated the need for the nested random effect modeling structure, and visual inspections of the models' diagnostic plots confirmed that model assumptions were met. 
Only the model for PC3 (Central $f_0$, Voice Clarity, Articulation Rate) of the ten PCs yielded a significant difference between speaker groups. This component comprises features of mean $f_0$ and $f_0$ minimum in semitones, mean Harmonics-to-Noise Ratio (HNR), and voiced segments per second, a rhythmic variable of measuring articulation rate. The LMM revealed a statistically significant effect ($\beta$ = 1.64, \textit{SE} = 0.49, \textit{df} = 62.04, \textit{t} = 3.37, FDR-adjusted \textit{p} = 0.01, 95\% CI [0.70, 2.58]), indicating that the autistic speakers of Slovak scored significantly lower on this component compared to the control speakers. Interpreting this direction, the finding suggests the autistic group's speech was characterized by a lower mean $f_0$, lower harmonicity (HNR), and/or a slower articulation rate than the control group's speech. 

This analysis aligns with the cross-linguistic findings regarding the central $f_0$; in both models, the control group scored significantly higher on the pitch-related component. However, meaningful divergences emerged regarding voice quality. While the cross-linguistic PC3 grouped pitch with intensity dynamics, the Slovak PC3 linked it with articulation rate and HNR, associating increased vocal clarity with the control group within this language. This contrasts with the cross-linguistic results, where a distinct voice quality component (PC4) associated higher harmonicity and clarity with the autistic group, highlighting that acoustic markers of voice quality may not be immutable traits of autism, but are influenced by the linguistic baselines of the population. These findings caution against a simplistic, unified description of an 'autistic voice,' while simultaneously emphasizing that voice quality remains a key, albeit context-dependent, differentiator across languages.

\section{Discussion}
The cross-linguistic analysis revealed three significant components that shed light on acoustic prosodic patterns distinguishing between autistic and non-autistic speakers. A robust finding was made for PC7 (Unvoiced Spectral Quality, Overall Intensity Consistency), where autistic speakers scored significantly higher, indicating that their speech was characterized by greater intensity variability when measured as coefficient of variation, providing further support for observations of varying intensity, which have lacked empirical clarity \cite{olivati2017acoustic, fusaroli2017voice}. The results suggest that a key differentiator may not be mean intensity, but variation, possibly linked to differences in maintaining consistent intensity modulation. Interestingly, in the analysis of PC3, increased dynamics of \textit{temporal slopes} of the intensity contour were associated with the control speakers. This separation of types of intensity variation highlights the need for methodological precision when characterizing variability. Rather than viewing intensity differences as a monolithic trait, our results reveal a heterogeneous profile where the autistic group exhibited greater overall fluctuation while the non-autistic group displayed steeper contours. This nuance emphasizes intensity variability as a multidimensional construct; resolving the conflicting findings in prior literature likely requires distinguishing between stability and dynamic modulation to fully capture the diverse prosodic features in autism.

The analysis of PC4 (Voice Clarity, Harmonicity) revealed that autistic speakers scored higher on a component defined by high HNR and spectral slope, corresponding to a clearer, less breathy voice quality. This finding challenges deficiency-centered models; in our cross-linguistic sample, phonation of autistic speakers was clearer than that of non-autistic peers, whose lower scores suggested a comparatively breathier phonation, providing empirical evidence regarding voice quality. Autistic speakers scored lower on PC3 (Intensity Dynamics, Central $f_0$, Spectral Change), indicating that their speech was characterized by lower central $f_0$, which adds to the heterogeneity of pitch findings in literature, where both elevated and reduced $f_0$ have been reported \cite{trayvick2024speech, fusaroli2017voice}.

Language-specific analyses provided further context. In the Finnish dataset, the only significant difference (PC2) mirrored the cross-linguistic voice quality findings: non-autistic speakers exhibited a breathier voice quality, while autistic speakers displayed a clearer, more modal voice. The Slovak analysis aligned with the cross-linguistic model regarding $f_0$, with autistic speakers exhibiting lower mean $f_0$. However, a divergence emerged in voice quality; within the Slovak model, higher HNR was linked to the control group. Acoustic markers of voice quality may be influenced by linguistic and cultural baselines of the population, cautioning against a simplistic, unified description of voice quality in autism. Still, the robust cross-linguistic findings of voice quality warrant further study.

This study provides relevant empirical findings by utilizing a robust methodology: using PCA to reduce 88 correlated features into orthogonal components, to accurately model complex acoustic patterns. Crucially, the strongest findings were not defined primarily by $f_0$, but by voice quality and intensity dynamics. This supports the call to broaden the focus of prosodic research in autism to less studied phenomena. 

The limitations of this study should be acknowledged, the first being the imbalanced nature of the data. The Slovak corpus outnumbered the Finnish and French datasets, meaning the cross-linguistic results are weighted by the Slovak data. In addition, while the nested random effects structure was designed to control for language-level baselines, the small sample size for French precluded robust monolingual modeling. Finnish and French datasets are also demographically quite homogeneous, including a narrower age range of only male speakers. Thus, while the patterns identified in this study are compelling, these results should be viewed as exploratory rather than definitive. Furthermore, the observed language-specific nuances have not yet been mapped in detail to the unique phonological structures of each language. Future cross-linguistic research should integrate phonological theory in the analysis of prosodic features and their interaction with other areas of language in autism.

Additionally, the Finnish data contained a slight situational confound: autistic participants were recorded in a naturalistic intervention session in a familiar group setting, whereas control participants were recorded in a constructed conversational setting, which may have influenced prosodic spontaneity independent of diagnosis. Finally, detailed clinical metadata regarding autism severity were not available for the corpora. However, all participants were verbal and able to complete collaborative tasks, which suggests a profile consistent with lower support needs, as do the (nowadays outdated) Asperger's syndrome diagnoses of the Finnish autistic speakers. 

Despite these limitations, the study offers value by identifying stable acoustic patterns, particularly the less-commonly reported voice quality features, across a diverse linguistic sample. The consistencies observed in the cross-linguistic model suggest that these prosodic markers may transcend specific language environments, providing a compelling foundation for future, more balanced cross-linguistic investigations. Future research should replicate this approach in larger, more balanced corpora to investigate the language-independence of the acoustic markers, and investigate their perceptual salience to listeners. 

\section{Conclusions}
This study identified robust prosodic differences between autistic and non-autistic speakers. While prosody of non-autistic speakers was associated with higher pitch and varying intensity slopes, prosody in autism was characterized mostly by a less breathy, modal voice quality and increased intensity variability. These results reinforce existing findings while introducing critical nuance by highlighting the role of voice quality and intensity in the analysis of cross-linguistic markers of prosody in autism.

\section{Acknowledgements}
Regarding funding, ILM was funded by the Research Council of Finland under grant number 362906, and SK was supported by the L-SCALE project funded by Kone Foundation. The authors wish to acknowledge CSC – IT Center for Science, Finland, for providing the computational resources.

\bibliographystyle{IEEEtran}

\bibliography{mybib}

% \begin{thebibliography}{9}
% \bibitem[1]{Davis80-COP}
%   S.\ B.\ Davis and P.\ Mermelstein,
%   ``Comparison of parametric representation for monosyllabic word recognition in continuously spoken sentences,''
%   \textit{IEEE Transactions on Acoustics, Speech and Signal Processing}, vol.~28, no.~4, pp.~357--366, 1980.
% \bibitem[2]{Rabiner89-ATO}
%   L.\ R.\ Rabiner,
%   ``A tutorial on hidden Markov models and selected applications in speech recognition,''
%   \textit{Proceedings of the IEEE}, vol.~77, no.~2, pp.~257-286, 1989.
% \bibitem[3]{Hastie09-TEO}
%   T.\ Hastie, R.\ Tibshirani, and J.\ Friedman,
%   \textit{The Elements of Statistical Learning -- Data Mining, Inference, and Prediction}.
%   New York: Springer, 2009.
% \bibitem[4]{YourName17-XXX}
%   F.\ Lastname1, F.\ Lastname2, and F.\ Lastname3,
%   ``Title of your INTERSPEECH 2021 publication,''
%   in \textit{Interspeech 2021 -- 20\textsuperscript{th} Annual Conference of the International Speech Communication Association, September 15-19, Graz, Austria, Proceedings, Proceedings}, 2020, pp.~100--104.
% \end{thebibliography}

\end{document}